\documentclass[aps,onecolumn,groupedaddress,11pt,tightenlines]{revtex4-1}
\usepackage{graphicx}
\usepackage{amssymb}
\usepackage{amsmath}
\usepackage[caption=false]{subfig}
\begin{document}

\title{Regular and Chaotic Dynamics of a Piecewise Smooth Bouncer\footnote{\textit{Preprint submitted to Chaos November 11, 2014}}}

\author{Cameron K. Langer}
\email{c.k.langer@tcu.edu}
\affiliation{Department of Physics and Astronomy, Texas Christian University}
\author{Bruce N. Miller}
\email{b.miller@tcu.edu}
\affiliation{Department of Physics and Astronomy, Texas Christian University}

\begin{abstract}
The dynamical properties of a particle in a gravitational field colliding with a rigid wall moving with piecewise constant velocity are studied. The linear nature of the wall's motion permits further analytical investigation than is possible for the system's sinusoidal counterpart. We consider three distinct approaches to modeling collisions: (i) elastic, (ii) inelastic with constant restitution coefficient and (iii) inelastic with a velocity-dependent restitution function. We confirm the existence of distinct unbounded orbits (Fermi acceleration) in the elastic model, and investigate regular and chaotic behavior in the inelastic cases. We also examine in the constant restitution model trajectories wherein the particle experiences an infinite number of collisions in a finite time i.e., the phenomenon of inelastic collapse. We address these so-called ``sticking solutions'' and their relation to both the overall dynamics and the phenomenon of self-reanimating chaos. Additionally, we investigate the long-term behavior of the system as a function of both initial conditions and parameter values. We find the non-smooth nature of the system produces novel bifurcation phenomena not seen in the sinusoidal model, including border-collision bifurcations. The analytical and numerical investigations reveal that although our piecewise linear bouncer is a simplified version of the sinusoidal model, the former not only captures essential features of the latter but also exhibits behavior unique to the discontinuous dynamics.
\end{abstract}

\maketitle

\section{\label{sec:1} Introduction}

The problem of a particle bouncing on a vertically and periodically driven wall has been studied extensively since a similar system was introduced by Fermi and Ulam \cite{fermi1949} in the study of high-energy cosmic rays. Originally, it was proposed by Fermi that a particle colliding elastically between one fixed and one oscillating wall could in theory obtain unbounded energy growth. Investigations by Pustyl'nikov in \cite{pustylnikov1968,pustylnikov1972,pustylnikov1983} showed that although in the Fermi-Ulam model energy remained bounded for sufficiently smooth wall motions, unbounded orbits could be found in the related system of a particle colliding with a sinusoidally oscillating wall in the presence of a constant gravitational field. This system was also found to exhibit a variety of complex and diverse dynamical behavior for both elastic and inelastic collisions, and has been widely studied (cf. for early studies \cite{lichtenberg1980,holmes1982,lichtenberg1992,luck1993} and \cite{naylor2002,vogel2011} for more recent investigations).

From a mathematical perspective, the gravitational bouncer and related Fermi-Ulam models with piecewise-smooth wall motions have been studied at large energies as perturbations of integrable systems; however, since these perturbations are not smooth, KAM theory can no longer be applied, and the normal (limiting) form of the map becomes more complicated \cite{dolgopyat2008,desomoi2012}. Understanding these simple models offer first glimpses into the general class of piecewise smooth, nearly integrable systems, an area of research where many open questions remain \cite{dolgopyat2013}.

While numerous numerical and experimental studies of this system have been done, the usual choice of a sinusoidal driving force prevents the explicit Poincar\'{e} map of this system from being analytically computed. Hence, there are certain restrictions which have to be made in any theoretical work on the sinusoidal model, and often approximate systems have been studied in place of the exact model. In this work we introduce the so-called \textit{piecewise linear bouncer}, consisting of a particle colliding with a wall which is being driven by a piecewise linear function (which makes the wall's \textit{velocity} piecewise constant). Choosing a piecewise linear $f(t)$ as the driving function allows the Poincar\'{e} surface-of-section or impact map $\mathcal{T}:(t_n,v_n)\rightarrow(t_{n+1},v_{n+1})$ to be analytically computed, making exact study of the system at hand possible. This allows us to proceed further analytically for this model than is possible in the sinusoidal model. However, the discontinuity present in the wall's motion introduces further complexity in the dynamics, leading to a Poincar\'{e} map which is only piecewise-smooth. Hence, we have replaced a smooth map for which the explicit form could not be computed analytically with a non-smooth map for which the explicit form of the map is known.

Non-smooth dynamical systems arise in a variety of physical applications (such as switching circuits), a prominent class of which is the so-called \textit{impact oscillator} (cf. \cite{diBernardo2008} for an introduction to such systems). One example of an impact oscillator is the widely studied motion of a particle attached via a linear spring and dashpot to a datum point such that the position is given by $\ddot{x}+2\zeta\dot{x}+x=w(t)$ for $x>f$, where $x$ is the particle's position, $2\zeta$ is a measure of the damping, and $w$ is an applied external force. At the point $x=f$ the particle collides with some rigid obstacle and follows the impact law $x^+=x^-$ and $v^+=-\epsilon v^-$, where $x^+,v^+$ and $x^-,v^-$ denote the position and velocity of the particle just before and after a collision, respectively. The equivalent problem of setting $w=0$ and allowing $f=f(t)$ to move and collide with the particle according to $v^+=(1+\epsilon )\dot{f}(t)-\epsilon v^-$, where $\epsilon$ is the restitution coefficient, shows the relation between the one-dimensional impact oscillator and our piecewise linear bouncer. Many of the features of impact oscillators, such as \textit{sticking regions} and \textit{inelastic collapse} are features present in our model as well. The latter feature, consisting of infinite collisions occurring in finite time, is also observed in studies of granular media \cite{mcnamara1992,goldman1998}. Additionally, some of the bifurcation phenomena unique to non-smooth systems are found in our model when the parameters of the system are varied, including \textit{period-adding}, discontinuous appearance/disappearance of periodic and chaotic orbits, and sudden transitions from periodic to chaotic orbits. All of these fall in the domain of \textit{discontinuity-induced bifurcations}, or DIBs for short (cf. \cite{nusse1992,banerjee1999,diBernardo1999} for an introduction to discontinuity-induced bifurcations).

We find the piecewise linear bouncer offers the opportunity to study in greater analytical detail some of the essential aspects found in the sinusoidal model, as well as some additional complexity due to the discontinuities in the wall's motion. In what follows we study the elastic model as well as two different kinds of inelastic models: a constant restitution coefficient $\epsilon\in (0,1)$ and a velocity-dependent restitution function $\epsilon=\epsilon(v_{n+1}^-)$ which is dependent on the velocity of the particle just before impact. We use the particular restitution function found in \cite{weir2005}, with $\epsilon$ dependent on the relative velocity of the particle, \begin{equation}\epsilon(v_{n+1}^-)=\exp\left\{-\Big\vert \frac{v_{n+1}^--\dot{f}}{c}\Big\vert^{3/5}\right\},\end{equation} where $c$ is a constant which makes the argument dimensionless, and $v_{n+1}^-$ is the velocity of the particle just before the $n+1^{\text{st}}$ collision, given by $v_{n+1}^- =-g(t_{n+1}-t_n)+v_n$.

We note that previous work has been done by Okninski et al. in \cite{okninski2009,okninski2009-2} on a similar inelastic, constant restitution coefficient bouncer \footnote{The magnitude of the speed of the wall was \textit{not} constant in this model. In their work Okninski et al. considered either a sawtooth wave or an asymmetrical triangle wave as a driving function. In our work we consider a symmetrical triangle wave.}. In our work we confirm some of the basic results found in these investigations, as well as analyze some more global aspects of the dynamics. In their work Okninski et al. were limited to solving particular cases of the implicit map; our explicit map provides the opportunity to investigate the global dynamics for wide ranges of initial conditions and parameter values.

In this work, we study the Fermi piston with piecewise linear and periodic driving function for elastic as well as two different kinds of inelastic collisions. In section II, we compute the time of the $n+1^\text{st}$ collision; combining this with the collision equation(s) relating pre- and post-collision velocities gives the full Poincar\'{e} map. Next, we solve for periodic motions (i.e. ``regular solutions'') for each of the different kinds of collisions, and give a general description of the more exotic possible behavior (i.e. quasiperiodic, sticking, chaotic, and unbounded) for each of the different kinds of collisions. In sections III-V we present our numerical results for the three models which give a more global picture of the dynamics, and illustrate the dominance of certain solutions. In addition we discuss interesting features of chaotic motions observed in the velocity-dependent model, as well as scans of the parameter space to show the distribution of solutions. Finally, in section VI we summarize our results and indicate future directions of research.

\section{\label{sec:2}Basics}
\subsection{\label{sec:2.1}The impact map}
We consider the motion of a particle moving in one dimension, subject to a gravitational acceleration $g$, bounded from below by an infinitely massive wall which is being driven by a function $f(t)$. We assume that collisions do not affect the wall's motion i.e. the wall is infinitely massive. Between the $n^\text{th}$ and $n+1^\text{st}$ collision, the particle is in free fall. Thus, after defining the wall's motion to be given by \begin{equation}f(t)=\left\{\begin{array}{cc} \frac{4At(\text{mod}\,T)}{T}-A, & t(\text{mod}\,T)<T/2 \\ -\frac{4At(\text{mod}\,T)}{T}+3A, & t(\text{mod}\,T)>T/2 \end{array}\right. ,\end{equation} the time-of-flight between the $n^\text{th}$ collision at $t_n$ and the $n+1^\text{st}$ collision at $t_{n+1}$ is given by the smallest positive solution of \begin{equation}\label{eq:1}-\frac{1}{2}g(t_{n+1}-t_n)^2+v_n(t_{n+1}-t_n)+f(t_n)-f(t_{n+1})=0,\end{equation} and the post-collision velocity of the particle is given by \begin{equation}\label{eq:2}v_{n+1}^+:=v_{n+1}=(1+\epsilon)\dot{f}(t_{n+1})+\epsilon g(t_{n+1}-t_n)-\epsilon v_n.\end{equation} To reduce the number of parameters of the system, we introduce the dimensionless acceleration $a:=gT^2/A$ and henceforth work in units where $T=A=1$. It is convenient in solving the above equation to define the cyclic time coordinate $\theta:=t\,\text{mod}\,T=t\,\text{mod}\,1$ and reduce the phase space to a half cylinder. We chose to distinguish the different possible kinds of collisions as follows. If the particle fails to escape the collision zone $[-A,A]=[-1,1]$ between collisions we say a \textit{direct} collision occurs; otherwise we say an \textit{indirect} collision occurs. By definition for direct collisions we have three possible cases for the motion of the wall at the $n^\text{th}$ and $n+1^\text{st}$ collisions: either the particle collides with a wall moving upward at both collisions, downward at both collisions, or down at the $n^\text{th}$ collision and up at the $n+1\text{st}$ collision. In the first two cases, the substitution $t_{n+1}-t_n=\theta_{n+1}-\theta_n$ may be made, while in the latter we must use $t_{n+1}-t_n=\theta_{n+1}-\theta_n+1$ because the wall completes an oscillation in the time between collisions. For indirect collisions, instead of computing $t_{n+1}-t_n$ directly we first compute $t_\text{fall}-t_n$, the time it takes the particle to re-enter the collision zone ($t_\text{fall}$ denotes the time of the particle's reentry). In terms of $t_\text{fall}$, the time of the next collision $t_{n+1}$ is given by \begin{equation}-\frac{a}{2}(t_{n+1}-t_\text{fall})^2+v_\text{fall}(t_{n+1}-t_\text{fall})+1-f(t_{n+1})=0.\end{equation} In solving this equation we distinguish once again three general kinds of collisions: the wall is either moving upward or downward at the time the particle re-enters the collision zone \textit{and} at the moment of the $n+1^\text{st}$ collision, \textit{or} the wall is moving downward upon re-entry and upward at the moment of collision. It follows that in the former two cases the substitutions $t_{n+1}-t_\text{fall}=\theta_{n+1}-\theta_\text{fall}$ may be made, and in the latter case $t_{n+1}-t_\text{fall}=\theta_{n+1}-\theta_\text{fall}+1$. Upon making these substitutions and solving the subsequent quadratic equations, we arrive at the map (where we take $\theta_{n+1}\,\text{mod}\,1$ in all cases) \par\footnotesize\begin{equation}\label{eq:3} \theta_{n+1}=\left\{\begin{array}{lr} \theta_n+{2}\left[v_n\mp4\right]/a, & v_n<\zeta,t_\text{pA}>t_\text{wA}\,\&\,\theta_n<1/2\,\text{or}\,t_\text{pF1}<t_\text{wF1} \\ \theta_n+\left[v_n-4\pm\sqrt{(v_n-4)^2+8a+2\delta_n}\right]/a, & v_n<\zeta,t_\text{pA}>t_\text{wA},\theta_n>1/2\,\&\,t_\text{pF1}>t_\text{wF1} \\ \theta_n+\left[v_n\mp 4+\sqrt{(\sqrt{v_n^2+\delta_n}\mp 4)^2+\delta_\text{fall}}\right]/a, & v_n>\zeta,t_\text{pA}<t_\text{wA},\theta_\text{fall}<1/2\,\text{or}\,t_\text{pF2}<t_\text{wF2} \\ \theta_n+\left[v_n-4+\sqrt{(\sqrt{v_n^2+\delta_n}-4)^2+8a+\delta_\text{fall}}\right]/a, & v_n>\zeta,t_\text{pA}<t_\text{wA},\theta_\text{fall}>1/2\,\&\,t_\text{pF2}>t_\text{wF2} \end{array}\right. ,\end{equation} \begin{equation}\label{eq:4} v_{n+1}=\left\{\begin{array}{lr} \epsilon\left[v_n\mp 4\right]\pm 4, & v_n<\zeta,t_\text{pA}>t_\text{wA}\,\&\,\theta_n<1/2\,\text{or}\,t_\text{pF1}<t_\text{wF1} \\ \epsilon\sqrt{(v_n-4)^2+8a+2\delta_n}+4+\epsilon a, & v_n<\zeta,t_\text{pA}>t_\text{wA},\theta_n>1/2\,\&\,t_\text{pF1}>t_\text{wF1} \\ \epsilon\sqrt{(\sqrt{v_n^2+\delta_n}\mp 4)^2+\delta_\text{fall}}\pm 4, & v_n>\zeta,t_\text{pA}<t_\text{wA},\theta_\text{fall}<1/2\,\text{or}\,t_\text{pF2}<t_\text{wF2} \\ \epsilon\sqrt{(\sqrt{v_n^2+\delta_n}-4)^2+8a+\delta_\text{fall}}+4+\epsilon a, & v_n>\zeta,t_\text{pA}<t_\text{wA},\theta_\text{fall}>1/2\,\&\,t_\text{pF2}>t_\text{wF2} \end{array}\right. ,\end{equation}\normalsize where we have defined the useful quantities \begin{equation}\delta_n:=\left\{\begin{array}{cc} -4a(1-2\theta_n), & \theta_n<1/2 \\ 4a(1-2\theta_n), & \theta_n > 1/2\end{array}\right. ,\end{equation} \begin{equation}\theta_\text{fall}:=t_\text{fall}\,\text{mod}\,1=\theta_n+\left[v_n+\sqrt{v_n^2+\delta_n}\right]/a\,\,\text{mod}\,1,\end{equation} and \begin{equation}\delta_\text{fall}:=4a(1-2\theta_\text{fall}).\end{equation} In the first case we take the $(-)$ sign if the wall is moving upward at the $n^\text{th}$ collision i.e., $\theta_n<\frac{1}{2}$ and vice versa for $\theta_n>\frac{1}{2}$. In the second case we include the $\pm$ in front of the square root to ensure the smallest \textit{positive} solution for the time interval is taken, and in the third case the $\mp$ and $\pm$ again depend on the wall's motion at the $n^\text{th}$ collision. For example if $\theta_n< \frac{1}{2}$ then we take the mapping to be \begin{equation}\theta_{n+1}=\theta_n+\frac{v_n-4+\sqrt{(\sqrt{v_n^2+\delta_n}-4)^2+\delta_\text{fall}}}{a}\,\,\text{mod}\,1,\end{equation} \begin{equation}v_{n+1}=\epsilon\sqrt{(\sqrt{v_n^2+\delta_n}-4)^2+\delta_\text{fall}}+4.\end{equation}  Note that if we neglect the modulo operator in $\theta_\text{fall}$, the third and fourth cases reduce to the first and second cases, respectively. 

By deriving the impact map~(\ref{eq:3})-(\ref{eq:4}) we have reduced the study of the piecewise linear bouncer to a piecewise-defined map (where $\mathbf{x}_n:=(\theta_n, v_n)$) \begin{equation}\mathbf{x}_{n+1}=\mathbf{f}_i(\mathbf{x}_n).\end{equation} In subsequent analysis we work with either the implicit map defined by~(\ref{eq:1})-(\ref{eq:2}) or the explicit map defined by~(\ref{eq:3})-(\ref{eq:4}). 

\subsection{\label{sec:2.2}Types of solutions}
\subsubsection{\label{sec:2.2.1}Elastic and constant restitution models - fixed and periodic solutions}
In general, working directly from~(\ref{eq:3})-(\ref{eq:4}) in searching for fixed points and periodic orbits we arrive at two simultaneous nonlinear equations which must be solved numerically for $(\theta_n,v_n)$ which involve (due to the presence of the modulo operator in the $\theta_{n+1}$ map) at least one arbitrary integer. In searching for basic periodic solutions, we have found it to be more convenient to work directly from~(\ref{eq:1})-(\ref{eq:2}), the implicit map. However, the two approaches are equivalent and any solutions derived from this map are also valid for the explicit map, if the right combination of the different cases is employed.

 We begin our search for fixed point solutions by supposing the particle impacts with an upward-moving wall; the condition for fixed point $\theta_{n+1}=\theta_n$ implies that $t_{n+1}-t_n=p\in\mathbb{Z}^+$. By imposing this condition along with $v_{n+1}=v_n=v_*$ on the implicit map we arrive at \begin{equation}-\frac{a}{2}p^2+v_*p=0,\end{equation} \begin{equation}v_*=4(1+\epsilon)+\epsilon a p-\epsilon v_*,\end{equation} which implies \begin{equation}v_*=\frac{ap}{2}=4+\frac{\epsilon a p}{1+\epsilon}=\frac{4(1+\epsilon)}{1-\epsilon}\,\,\iff\,\,a=\frac{8(1+\epsilon)}{p(1-\epsilon)}.\end{equation} Hence we find the \textit{family} of fixed points \begin{equation}\label{eq:5} \theta_*\in(0,\frac{1}{2}),\,\,\,v_*=\frac{4(1+\epsilon)}{1-\epsilon},\,\,\,\,\text{valid\,for}\,a=\frac{8(1+\epsilon)}{p(1-\epsilon)},\,\,p=1,2,\ldots ,\end{equation} where $\theta_*$ is some arbitrary value in the range $(0,\frac{1}{2})$. Note that for the elastic case $\epsilon=1$ the only fixed point corresponds to the particle matching the wall's velocity exactly.

\begin{figure}[h]
\subfloat[Subfigure 1 list of figures text][]{\includegraphics[height=5.5cm, width=7.5cm]{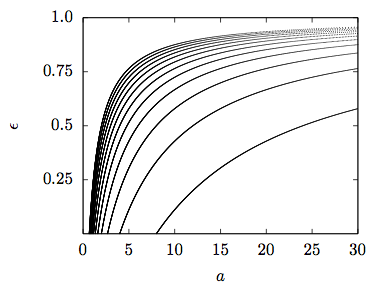}
\label{fig:arnold}}
\qquad
\subfloat[Subfigure 2 list of figures text][]{\includegraphics[height=5.5cm, width=7.5cm]{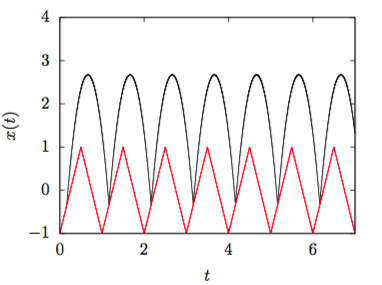}
\label{fig:realtime}}
\caption{(a) The first ten fixed point solutions in parameter $a,\epsilon$-space. (b) A real-time fixed point trajectory.}
\label{fig:one}
\end{figure}

 Although there exist analogous fixed points for the sinusoidal model where the flight time equals an integer multiple of the wall's period of oscillation, the nature of the fixed points are fundamentally different for our model. In our case the solutions only exist for sharply defined values of the parameter $a$, while in the sinusoidal model the solutions exist and are stable for a region of parameter space. Thus for the sinusoidally driven model the solutions correspond to resonances wherein the `period' of the particle's flight between two collisions is an integer $p$ times larger than the period of the wall's motion. The regions in parameter space where these resonances exist define a collection of the so-called \textit{Arnol'd tongues}. In our case, since the solutions exist only for sharply defined parameter values, the collection of fixed points defines a set of isolated curves in the parameter space e.g., Arnol'd tongues of zero width everywhere, as displayed in Fig.~\ref{fig:arnold}. 

Performing linear stability analysis on this orbit requires finding the eigenvalues of $D\mathbf{f}$, the Jacobian matrix \begin{equation}D\mathbf{f}=\left[\begin{array}{cc} \partial \theta_{n+1}/\partial \theta_n & \partial \theta_{n+1}/\partial v_n \\ \partial v_{n+1}/\partial \theta_n & \partial v_{n+1}/\partial v_n \end{array}\right].\end{equation} For this orbit a simple calculation leads to the eigenvalues $\lambda=1$ and $\lambda=\epsilon^2<1$. Thus, this orbit is \textit{marginally stable}, since if we perturb a fixed point in the $\theta$ eigendirection the orbit does not return to this value. However, in the sense that the orbit remains a part of the family of fixed points, we may say that the manifold formed by the collection of fixed points at a given $p$ is in fact \textit{stable}, since a perturbed point on the manifold of fixed points will stay on that manifold.

Due to the inherent ambiguity present in the cyclic variable $\theta$, any search for a period-$2$ orbit i.e., an orbit with $\theta_{n+2}=\theta_n$ and $v_{n+2}=v_n$ will involve an arbitrary integer corresponding to the number of wall oscillations between $t_{n+1}$ and $t_n$. In general, we may classify all periodic orbits as follows: 

\begin{itemize}
\item Period-$q^\text{up/down}$ orbits, where all $q$ collisions in the orbit are with a wall moving in the same direction. These orbits are similar in nature to the fixed points, and result in a free variable in $\theta_n$.
\item Period-$q^{m,n}$ orbits, where $m$ collisions occur with an upward-moving wall, and $n$ collisions occur with a downward moving wall, with $q=m+n$. These orbits are more similar to periodic solutions observed in the sinusoidal model.
\end{itemize}

In this work, we limit our investigations to period-$2$ orbits, and derive the general constraint equations for both of the possible periodic orbits. 

If a period-$2$ orbit consists of an initial collision with an upward-moving wall followed by a second collision with a downward-moving wall i.e., $\theta_n<\frac{1}{2}$ and $\theta_{n+1}>\frac{1}{2}$, the implicit mapping reads (where we have used $t_{n+2}=t_n+p$ for $p\in\mathbb{Z}^+$) \begin{equation}-\frac{a}{2}(t_{n+1}-t_n)^2+v_n(t_{n+1}-t_n)+4[\theta_n+\theta_{n+1}-1]=0,\end{equation} \begin{equation}-\frac{a}{2}(t_n-t_{n+1}+p)^2+[-4(1+\epsilon)+\epsilon a (t_{n+1}-t_n)-\epsilon v_n](t_n-t_{n+1}+p)-4[\theta_n+\theta_{n+1}-1]=0,\end{equation} \begin{equation}v_n=4(1+\epsilon)+\epsilon a (t_n-t_{n+1}+p)-\epsilon[-4(1+\epsilon)+\epsilon a (t_{n+1}-t_n)-\epsilon v_n].\end{equation} We write $t_n=\theta_n+k_n$, $t_{n+1}=\theta_{n+1}+k_{n+1}$ and define $\Delta \theta_{n+1}:=\theta_{n+1}-\theta_n$ and $k:=k_{n+1}-k_n$ where $k_n,k_{n+1}\in\mathbb{Z}^+$; then the above equations may be rewritten as \begin{equation}-\frac{a}{2}(\Delta\theta_{n+1}+k)^2+v_n(\Delta\theta_{n+1}+k)+4[2\theta_n+\Delta\theta_{n+1}-1]=0,\end{equation} \begin{equation}-\frac{a}{2}(p-k-\Delta\theta_{n+1})^2+[-4(1+\epsilon)+\epsilon a (\Delta\theta_{n+1}+k)-\epsilon v_n](p-\Delta\theta_{n+1}-k)-4[2\theta_n+\Delta\theta_{n+1}-1]=0,\end{equation} \begin{equation}(1-\epsilon^2)v_n=4(1+\epsilon)+\epsilon a (p-\Delta\theta_{n+1}-k)+4\epsilon(1+\epsilon)-\epsilon^2a(\Delta\theta_{n+1}+k).\end{equation} Eliminating $\Delta\theta_{n+1}$ yields the two coupled nonlinear equations \begin{multline}\label{eq:6}apv_n+\Big[\left(ap-v_n-4\right)\sqrt{(v_n+4)^2-8a(1+k-2\theta_n)}-\big[(v_n+4)^2 -8a(1+k-2\theta_n)\big]\Big]\left(\epsilon+1\right)=0,\end{multline} \begin{equation}\label{eq:7}\epsilon^2\left[(v_n+4)^2-8a(1+k-2\theta_n)\right]=\left[\frac{a\epsilon p+4(1+\epsilon)}{\epsilon+1}-v_n\right]^2.\end{equation} 

The general equations~(\ref{eq:6})-(\ref{eq:7}) constraining period-$2$ orbits where the collisions are with an upward/downward moving wall yield in theory an infinite number of solutions with respect to the integers $p$ and $k$ and the parameters $a$ and $\epsilon$. 

In addition to these period-$2^\text{1,1}$ orbits, we also have period-$2^\text{up/down}$ orbits. These cases are analytically simpler than the above calculations. Using the same notation as above, we have from the implicit map for consecutive collisions with an upward-moving wall \begin{equation}-\frac{a}{2}(\Delta\theta_{n+1}+k)^2+v_n(\Delta \theta_{n+1}+k)-4\Delta\theta_{n+1}=0,\end{equation} \begin{equation}-\frac{a}{2}(p-\Delta\theta_{n+1}-k)^2+[4(1+\epsilon)+\epsilon a(\Delta\theta_{n+1}+k)-\epsilon v_n](p-\Delta\theta_{n+1}-k)+4\Delta\theta_{n+1}=0,\end{equation} \begin{equation}v_n=4(1+\epsilon)+\epsilon a(p-\Delta\theta_{n+1}-k)-\epsilon[4(1+\epsilon)+\epsilon a (\Delta\theta_{n+1}+k)-\epsilon v_n].\end{equation} Eliminating $\Delta\theta_{n+1}$ yields the coupled equations \begin{equation}\label{eq:8}{\left(v_n-\frac{4\left(\epsilon+1\right)+ap\epsilon}{\epsilon+1}\right) }^{2}={\epsilon}^{2}\,\left( {\left( v_n-4\right) }^{2}+8ak\right),\end{equation} \begin{equation}\label{eq:9}2(\epsilon+1)(v_n-ap-8)+ap(ap-2v_n)+2(\epsilon+1)[(v_n-4)^2+8ak]=0.\end{equation} Note that just like in the case of fixed points, the constraining equations are free of any $\theta_n$ dependence. By solving the first equation for $v_n$ and substituting the result into the second equation we arrive at a general equation constraining the four quantities $a,\epsilon,k$ and $p$. First we examine the elastic case. Applying $\epsilon=1$ to the above results in a simple formula \begin{equation}v_n=\frac{ap}{4}+4-\frac{8k}{p},\end{equation} along with the constraint equation \begin{multline} ap\left[\frac{3ap}{4}-11\right]+\frac{k}{p}\left[\frac{256k}{p}-32\right]+16\left[2ak-1\right]=0\\ \implies\,a=\frac{22p-64k\pm2\sqrt{169p^2-608kp+256k^2}}{3p^2}.\end{multline} Let us remind the reader that $k$ is the number of oscillations the wall completes between the $n^\text{th}$ and $n+1^\text{st}$ impacts, and that $p$ is the number of oscillations the wall completes between the $n^\text{th}$ and $n+2^\text{nd}$ collisions. Hence, we have generated for the elastic model a set of families of period-$2$ orbits: \begin{equation}\theta_n\in(0,\frac{1}{2}), v_n=\frac{ap}{4}+4-\frac{8k}{p},\,\,\text{where}\,\,a\,\,\text{solves}\,\,(33).\end{equation} In order for the solutions to be physically meaningful, we must have $\theta_n,\theta_{n+1}<\frac{1}{2}$ and $v_n,v_{n+1}>\dot{f}(\theta_n)$; this places restrictions on the values of $p$ and $k$. Stability analysis results in a Jacobian with eigenvalues \begin{equation}\lambda_1=1,\,\,\,\lambda_2=\frac{(ap^2-32k)^2}{(ap^2+32k)\sqrt{a^2p^4+128ap^3-64akp^2+1024k^2}},\end{equation} and if we plug in the expression for $a$ given above the second eigenvalue becomes a complicated function of $k$ and $p$, the general form of which is unlikely to yield much insight. However, for the simple cases $p=1$, $k=0$ the function simplifies to give $\lambda_2=\frac{1}{3}$, and thus we have another marginally stable set of periodic orbits. 

Returning to the inelastic model, solving~(\ref{eq:8})-(\ref{eq:9})  in general leads to a complicated expression for $v_n$ and a constraint equation involving $a,\epsilon,p$ and $k$. Proceeding in this fashion is difficult; if we instead choose particular values of $p$ and $k$, we may derive a formula for $v_n$ as well as a relationship between $a$ and $\epsilon$ for which period-$2$ orbits exist. For example, by choosing $p=1$, we must have $k=0$, in which case the first of the above equations may be solved: \begin{equation}v_n=\frac{a(\epsilon^2-4)}{(\epsilon-1)(\epsilon+1)^2}+4,\end{equation} and substituting this relation into the second equation leads to \begin{equation}a=\frac{32}{1-\epsilon}-\frac{16}{1+\epsilon^2}-8.\end{equation} Hence, we have the family of period-$2$ orbits given by \begin{equation}\label{eq:10}\theta_n\in(0,1/2),\,\,\, v_n=\frac{4(\epsilon-3)(\epsilon^3-\epsilon^2-3\epsilon-3)}{(\epsilon-1)^2(\epsilon^2+1)},\,\,\,\text{for}\,\,a=\frac{32}{1-\epsilon}-\frac{16}{1+\epsilon^2}-8. \end{equation} Analyzing the stability of this solution leads to a Jacobian matrix which is upper triangular; thus the eigenvalues are the diagonal entries: \begin{equation}\lambda_1=1,\,\,\,\,\lambda_2=2\left[\epsilon^2-\frac{\surd{2}\epsilon^5}{\sqrt{8\epsilon^4+a(\epsilon-1)^2}}\right]=2\epsilon^2-\frac{\epsilon^5\sqrt{\epsilon^2+1}}{\sqrt{\epsilon^6-2\epsilon^3+2\epsilon+1}}.\end{equation} It may be shown that the condition for bifurcation i.e., $\lambda_2=1$ implies that $\epsilon=\epsilon^*\simeq 0.829146$ (given by a root of a $10^\text{th}$ order polynomial). Hence, for $0<\epsilon<\epsilon^*$ we have a set of marginally stable period-$2$ orbits, and for $\epsilon^*<\epsilon<1$ the family of period-$2$ orbits is unstable. 
 
We have taken our analytical computations further to the case of period-$3$ orbits, and the analysis results in similar, albeit more complicated constraining equations. Nevertheless, the prevalence of periodic orbits in the system is undoubtedly clear. 

\subsubsection{\label{sec:2.2.2}Elastic - unbounded solutions}
Unsurprisingly, we find for the elastic model the existence of trajectories with unbounded velocity i.e., the presence of \textit{Fermi acceleration}. Interestingly, we find that there are different kinds of `Fermi accelerating' orbits, including orbits where the velocity growth is chaotic and orbits where the growth is extremely structured and almost quasiperiodic. These orbits will be discussed in greater detail in section~\ref{sec:3}.

\subsubsection{\label{sec:2.2.3}Inelastic constant restitution - sticking solutions}
For the inelastic model we find the dynamics to be greatly influenced by the \textit{sticking solutions}, wherein the particle experiences an infinite number of collisions with the wall in a finite time, sticks to the wall and is released at the moment of discontinuity in the wall's motion. We call a solution \textit{sticking} if at some point in time in its trajectory the particle undergoes the \textit{inelastic collapse} process. In general, this kind of motion is typical of certain non-smooth dynamical systems; for instance, sticking arises in many models which include friction of some kind \cite{diBernardo2008}. In addition, inelastic collapse arises in the study of granular media and gases of inelastic particles \cite{bernu1990,mcnamara1992,goldman1998}. For our system, we were able to analytically determine the time required for inelastic collapse to occur (by transforming to the wall's reference frame, since for small velocities the mapping simplifies greatly), as well as limiting velocities below which the particle was guaranteed to stick to the wall after `complete' inelastic collapse. We compute the sum of the flight-times to be given by, for an initial condition $(\theta_n,v_n)$, \begin{multline}\sum_{k=1}^\infty \Delta t_k=\frac{2(v_n\mp 4)}{a}+\frac{2(v_{n+1}\mp 4)}{a}\cdots=\frac{2}{a}\Big[1+\epsilon+\epsilon^2+\cdots\Big]=\frac{2(v_n\mp 4)}{a}\sum_{k=0}^\infty \epsilon^k \\=\frac{2(v_n\mp4)}{a(1-\epsilon)}<\left\{\begin{array}{lr}\frac{1}{2}-\theta_n, & \theta_n<\frac{1}{2} \\ 1-\theta_n, & \theta_n>\frac{1}{2} \end{array}\right. ,\end{multline} (where the inequality arises to ensure the particle \textit{completes} inelastic collapse before the wall changes its direction of motion) from which we find the limiting velocities \begin{equation}\xi_{\text{up}}=\frac{a(1-\epsilon)(\frac{1}{2}-\theta_n)}{2}+4,\end{equation} \begin{equation}\xi_{\text{down}}=\frac{a(1-\epsilon)(1-\theta_n)}{2}-4,\end{equation} below which inelastic collapse is guaranteed to occur (where we have used $\xi_{\text{up}}$ to denote the limiting velocity for inelastic collapse with an upward-moving wall, and $\xi_{\text{down}}$ for a downward-moving wall). After a particle completes the sticking process, it is released from the wall with new `initial' conditions $(\theta_\infty,v_\infty)=(\frac{1}{2},4)$. This holds for both upward and downward inelastic collapse, as in the latter case once the wall changes its motion discontinuously at $\theta=1$ the particle immediately has velocity less than that of the upward moving wall, and thus enters a `grazing' motion with the wall until $\theta=\frac{1}{2}$, at which point it is `released' with velocity equal to that of the wall. 

These results together with the mapping allows us in section~\ref{sec:4} to understand the behavior of sticking solutions for certain parameter values. More precisely, because of the generic nature of the sticking process and the uniformity of the initial/final conditions, we are able to in some cases determine the precise periodic nature of sticking solutions. In this case, we define the \textit{periodicity of sticking} to mean the number of flight time intervals a particle completes between sticking processes. For example, if a particle experiences inelastic collapse and is released with the initial conditions $(\theta_\infty,v_\infty)=(\frac{1}{2},4)$, and after re-colliding with the wall for the first time possesses a velocity below the limiting velocity $\xi_{\text{up/down}}$, it must re-start the process of inelastic collapse. Since the particle has cycled through a single flight time between the initial conditions $(\frac{1}{2},4)$ and the first post-collapse collision $(\theta_1,v_1)$, we call an orbit such as this \textit{periodic sticking with period $1$}. 

Thus, the uniformity of sticking enables us to define a notion of periodicity for the sticking orbits, and for small enough values of $\epsilon$ it may be shown that the sticking must occur with period 1, and for some other range of values with period 2, and so on. For this model, we found all sticking solutions to be periodic. This is in contrast to results found for the sinusoidal model, where sticking solutions which are aperiodic have been observed numerically and gives rise to the phenomenon of \textit{self-reanimating chaos}. Our results, which find no occurrences of self-reanimating chaos, give credence to the idea that the self-reanimating chaos observed in the sinusoidal model may be a result of numerical errors, and not inherent to the true system \cite{vogel2011}. 

\subsubsection{\label{sec:2.2.4}Inelastic constant restitution - chaotic solutions}
We find that sticking solutions regulate the phase space. More precisely, any solutions which are chaotic eventually enter the so-called \textit{locking region} which `locks' the particle into the inelastic collapse process, and forces the orbit to become periodic. This prevents the realization of long-term trajectories which are aperiodic and non-sticking. 

\subsubsection{\label{sec:2.2.5}Inelastic velocity-dependent model - periodic}
For the velocity-dependent model, we begin our analysis by searching for fixed points where the wall is moving upward at the moment of collision and completes $p$ periods of oscillation between two collisions i.e., $t_{n+1}-t_n=p$. Then the first mapping equation, which does not depend on the nature of the restitution function, is equivalent to \begin{equation}\Delta t_{n+1}=\frac{2v_n}{a}\,\,\,\iff \,\,\,v_n=\frac{ap}{2}.\end{equation} The velocity map may be simplified by noting that $v_{n+1}^-=-a\Delta t_{n+1}+v_n=-v_n$ which implies \begin{equation}\epsilon(v_{n+1}^-)=\exp{\left\{-\Big\vert \frac{v_{n+1}^--\dot{f}}{c}\Big\vert^{3/5}\right\}}=\exp{\left\{-\Big\vert \frac{-v_n-4}{c}\Big\vert^{3/5}\right\}},\end{equation} and since for an upward wall the velocity is always $\geq 4$ we may for $c>0$ drop the absolute values signs, allowing us to re-write the velocity map $v_{n+1}=(1+\epsilon(v_{n+1}^-))\dot{f}-\epsilon(v_{n+1}^-)v_{n+1}^-$ as \begin{equation}v_n=\frac{4(1+\exp{[(v_n+4)/c]^{3/5}})}{1-\exp{[(v_n+4)/c]^{3/5}}}=4\coth\left\{\frac{1}{2}\left(\frac{v_n+4}{c}\right)^{3/5}\right\}.\end{equation} Setting the two expressions for $v_n$ equal yields \begin{equation}v_n=2^{5/3}c\times\left[\coth^{-1}{\left\{\frac{ap}{8}\right\}}\right]^{5/3}-4.\end{equation} Thus we again arrive at a family of fixed point solutions \begin{equation}\theta_*\in(0,1/2),\,v_*=2^{5/3}c\times\left[\coth^{-1}{\left\{\frac{ap}{8}\right\}}\right]^{5/3}-4,\,p=1,2,\ldots,\end{equation} where $\theta_*$ is arbitrary. Since the inverse hyperbolic cotangent function $\coth^{-1}(x)$ is real-valued for positive $x$ only when $x>1$, we see that these fixed point solutions become invalid when $ap\leq8$ and thus there are no fixed point solutions for the range \begin{equation}0<a\leq\frac{8}{p},\,\,\,\,\,\,p=1,2,3,\ldots .\end{equation}

 Stability analysis results in a Jacobian with $\lambda_1=1$ and $\lambda_2<1$ regardless of parameter values. However, due to the fact that we have a \textit{family} of fixed points for all $\theta_*\in(0,\frac{1}{2})$, and the orbit is stable in the $v$ direction, we once again consider the family of fixed points to be stable. In general analytic expressions for periodic orbits are unlikely, due to the transcendental equations which arise in the constraint equations.
 
\subsubsection{\label{sec:2.2.6}Inelastic velocity-dependent model - chaotic solutions}
By introducing our particular choice of restitution function, we eliminate the possibility of inelastic collapse. More precisely, we may in fact show that the time required for an infinite number of collisions must itself be infinite, which means that sticking solutions play no part in our velocity-dependent model. The sum of an infinite number of direct collisions with an upward-moving wall leads to \begin{multline}\sum_{i=1}^{\infty} \Delta t_i=\frac{2}{a}(v_0-4)\left[1+\sum_{n=1}^\infty \prod_{j=1}^n \epsilon_j\right]  =\frac{2}{a}(v_0-4)\left[1+\sum_{n=1}^\infty \prod_{j=1}^n \exp\left\{-\left\vert\frac{4-v_j}{c}\right\vert^{3/5}\right\}\right] .\end{multline} Hence, the convergence/divergence of the sum of the flight times hinges on the convergence/divergence of the sum-of-products\begin{equation}S\equiv \sum_{n=1}^\infty \prod_{j=1}^n \exp\left\{-\left\vert \frac{4-v_{j}}{c}\right\vert^{3/5}\right\}.\end{equation} This sum can only converge if the terms in the series tend to zero as $n\rightarrow\infty$, but the process of decaying collisions ensures that $\lim_{n\rightarrow\infty} v_n=4=\dot{f}$ and thus the product $\prod_{j=1}^n a_j\rightarrow 1$ as $n\rightarrow \infty$, which means the sum diverges, and it requires an infinite amount of time for an infinite number of collisions to occur. 

The lack of sticking solutions allow more `interesting' long-term trajectories which are aperiodic to exist in the velocity-dependent model. In fact, we find attracting aperiodic orbits which have positive leading Lyapunov exponent as well as riddled basins of attraction; these chaotic orbits are discussed more fully in section~\ref{sec:5.2}. 

\subsection{\label{sec:2.3}Trapping region}
Just as in the sinusoidal model, we are able to define a trapping region which limits the long-term behavior of the system. It is straightforward to show that for $\epsilon<1$ the long-term velocity of the particle is bound inside the \textit{trapping region} of phase space \begin{equation}\dot{f}(t)\leq v_{n+1}(t)\leq v_{\text{upper}}=\frac{1+3\epsilon}{1-\epsilon}\frac{4A}{T}=4\times\frac{1+3\epsilon}{1-\epsilon}.\end{equation}
For the velocity-dependent model, a similar trapping region may be analytically demonstrated. More precisely, the upper bound for long-term velocity becomes in the velocity-dependent model  \begin{equation}v_{\text{upper}}:= \frac{1+3\epsilon(v_{n+1}^-)}{1-\epsilon(v_{n+1}^-)}4=4\left[\frac{1+3\exp{\left\{-(8/c)^{3/5}\right\}}}{1-\exp{\left\{-(8/c)^{3/5}\right\}}}\right]=4\left[2\coth{\left\{\frac{2^{4/5}}{c^{3/5}}\right\}}-1\right],\end{equation} where the final equality holds for positive $c$. Hence the particle's phase portrait for the velocity-dependent model is contained in \begin{equation}\mathcal{M}:= \left\{ (\theta_n,v_n): 0<\theta_n<1, \dot{f}(\theta_n)\leq v_n\leq 4\left[2\coth{\left\{\frac{2^{4/5}}{c^{3/5}}\right\}}-1\right]\right\}\,.\end{equation} 

\section{\label{sec:3}Elastic model}
\begin{figure}[h]
\subfloat[Subfigure 1 list of figures text][]{\includegraphics[scale=0.85]{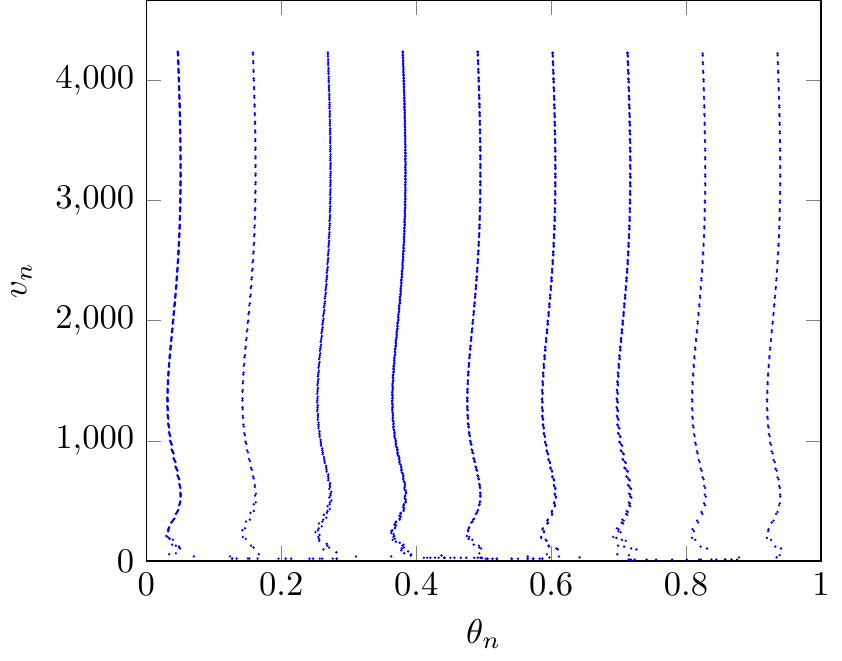}
\label{fig:elastic_one}}
\qquad
\subfloat[Subfigure 2 list of figures text][]{\includegraphics[scale=0.85]{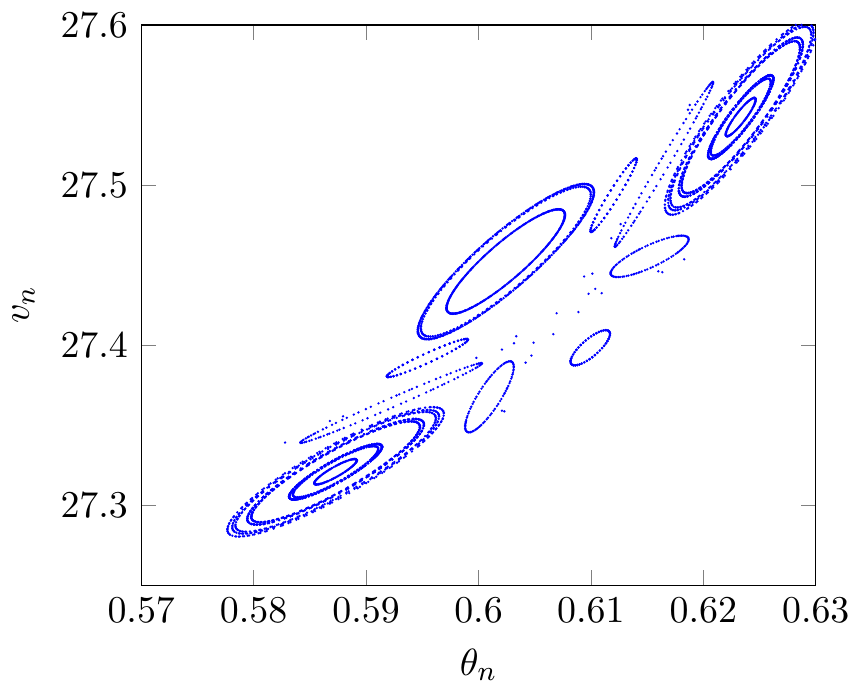}
\label{fig:elastic_two}}
\qquad
\subfloat[Subfigure 3 list of figures text][]{\includegraphics[scale=0.85]{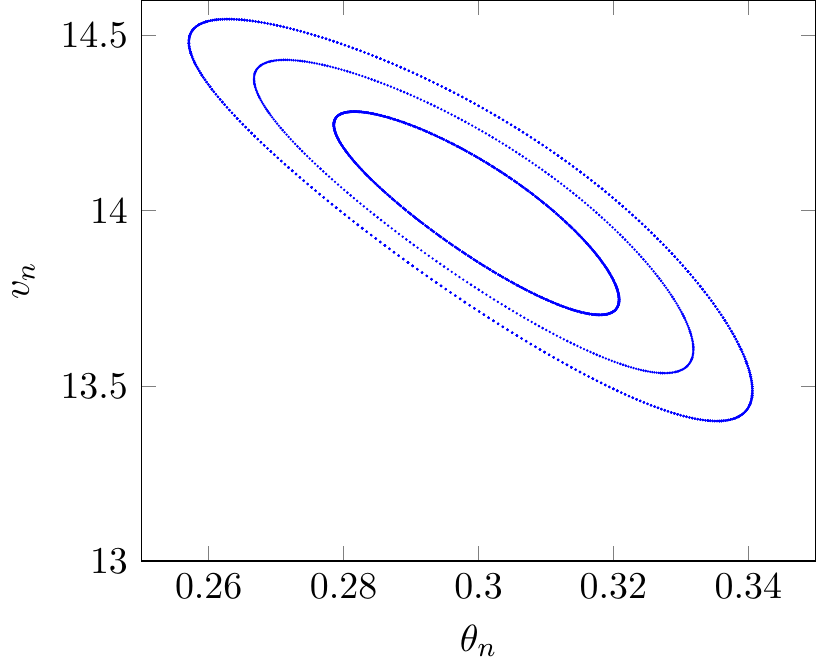}
\label{fig:elastic_three}}
\caption{Typical trajectories for the elastic model. (a) Branched Fermi acceleration for $a=19.0$; for this parameter value the trajectory is typical. (b) Quasiperiodic behavior resulting from a collection of initial conditions at $a=26.0$. (c) Enlarged view of an island of stability for $a=20.0$.}
\label{fig:elastic}
\end{figure}

For the elastic model, we find unbounded orbits which exist for different parameter values in addition to the usual periodic and quasiperiodic behavior. More precisely, we find \textit{several} kinds of orbits which lead to unbounded velocities, which are displayed in Fig.~\ref{fig:elastic}. These may be roughly classified as follows. First, we observe a structured behavior we call \textit{branched} Fermi acceleration. In this orbit, the particle's phase is `almost' periodic while the velocity is always increasing. This results in a trajectory which, upon plotting in a polar form with $\theta_n$ scaled by $2\pi$, forms a set of nine branches which grow outward as time moves forward and remain roughly at the same angles i.e., a set of nine angles $\phi_n:=2\pi \theta_n\simeq \text{constant}$. The existence of these branched accelerating orbits may be understood by comparing our map with the asymptotic form for large velocities. For this, the difference in position of the wall may be neglected, leading to a map of the form \begin{equation}\theta_{n+1}\simeq\theta_n+\frac{2[v_n-\dot{f}(\theta_{n+1})]}{a}\,\,\text{mod}\,1,\end{equation} \begin{equation}v_{n+1}\simeq v_n+\dot{f}(\theta_{n+1}),\end{equation} which naturally leads to unbounded orbits which are \textit{exactly} periodic in $\theta$, and `straighten out' the branches. Hence, the exact model provides slight corrections which are due to the wall's motion. 

In addition to this structured form of Fermi acceleration, we observe chaotic trajectories where the velocity growth is still unbounded, although the process is more unpredictable and slower. In these \textit{chaotic accelerating} orbits, the long-term trajectory takes the form of a chaotic sea which, if the surface-of-section is plotted in real-time, expands upward in the direction of increasing velocity, all the while avoiding the areas which form quasiperiodic islands of stability. These islands surround the other bounded orbits i.e., the period-$k$ solutions; these may be computed analytically for small $k$, and the corresponding quasiperiodic solutions arise in the usual fashion as evidenced by an irrational winding number. The examples displayed in Fig.~\ref{fig:elastic} illustrate the sensitivity of these orbits on the discontinuities of the wall's motion. If a periodic or quasiperiodic orbit hits a discontinuity, it is destroyed and the resulting orbit is chaotic and accelerating. Note that it is the absence of KAM invariant curves which (in the Fermi-Ulam model) span the phase space which allows for the unbounded velocity growth. 

\section{\label{sec:4}Constant coefficient of restitution}
\subsection{\label{sec:4.1}Typical trajectories}
For the inelastic model, we lose all instances of unbounded velocity growth. This is of course due to the presence of dissipation, which implies via Liouville's theorem that areas in the phase space contract; hence, for $\epsilon<1$ we should observe attractors, either regular or chaotic. In fact, we find that the existence of sticking solutions prevents any long-term chaotic trajectories from lasting beyond the first $\sim5\,000$ collisions. Thus, we observe two distinct behaviors: 

\begin{itemize}
\item Periodic attractors, valid only for specific parameter ranges;
\item Sticking attractors, which are periodic as well.
\end{itemize}
All numerical results lead to sticking solutions which repeat after a finite number of collisions. Hence, for the inelastic, constant restitution model the system's behavior is in all cases \textit{periodic}. 
\subsection{Bifurcation diagrams and typical trajectories}
\begin{figure}[h]
\subfloat[Subfigure 1 list of figures text][]{\includegraphics[height=5.5cm, width=7.5cm]{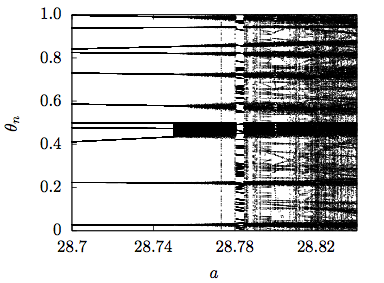}
\label{fig:bifurc_one}}
\qquad
\subfloat[Subfigure 2 list of figures text][]{\includegraphics[height=5.5cm, width=7.5cm]{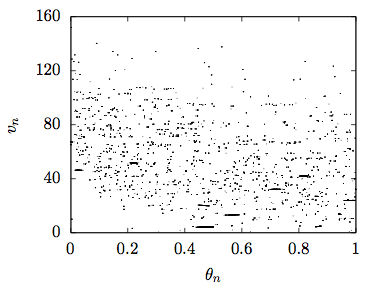}
\label{fig:sticking}}
\caption{(a) Portion of a bifurcation diagram for $\epsilon=0.96$, showing periodic, sticking, and apparently chaotic behavior. (b) An apparently chaotic solution at $a=28.82$ which is actually a sticking solution of period-$k>1000$. This is a generic feature of the constant restitution system, where motions which at first glance appear random possess finite periods of motion.}
\label{fig:bifurcation}
\end{figure}

In this model, large-scale bifurcation diagrams illustrating the evolution of individual trajectories as a single parameter is varied do not provide much insight, as sticking solutions which appear aperiodic may falsely lead one to believe the behavior is chaotic. However, if we enlarge individual regions of the diagrams and examine the fates of periodic orbits, we see that the destruction of these orbits is often due to points associated with the discontinuities in the wall's motion. In Fig.~\ref{fig:bifurcation} we display the typical behavior at high values of the restitution coefficient. The diagram shows a complicated mixing of periodic and sticking behavior. To generate these bifurcation diagrams, a particular initial condition $(\theta_0,v_0)$ was followed for $5\,000$ collisions, with only the final $500$ collisions displayed to eliminate transients, for a fixed $\epsilon$. Then the parameter $a$ was incremented, and the process repeated. 

Note that, since the inelastic collapse process takes an infinite number of collisions to occur, we have in our simulations used the limiting velocity $\xi_\text{up/down}$ to introduce a `cutoff.' If after iterating the map we find the velocity of the particle to be below this limiting velocity, we immediately skip the inelastic collapse process and set the phase and velocity to their post-sticking values i.e., $\theta_\infty=\frac{1}{2}$ and $v_\infty=4$. Hence in Fig.~\ref{fig:bifurcation}, the orbits with points lying on the horizontal line at $\theta=\frac{1}{2}$ correspond to sticking solutions. This allows us to see clearly how many collisions it takes the particle to re-start the sticking process i.e., the periodicity of sticking. When viewed on a large scale, the resulting orbits in Fig.~\ref{fig:bifurcation} appear to be chaotic. However, a closer examination on a smaller window of $a$ values reveals that the complex chaotic-looking orbit is in fact a sticking orbit with a very high period of sticking. Thus, we must take care in distinguishing the truly chaotic non-sticking orbits and the chaotic-looking but periodic sticking orbits when characterizing the system's behavior. One way to confirm this is by plotting a larger number of collisions; as the number of collisions goes up, a truly chaotic sea will look more and more complex, while a periodic sticking solution will at some point reach a point at which the orbit repeats and the surface-of-section stops changing.

\subsection{\label{sec:4.2}Basins of attraction}
\begin{figure}
\includegraphics[height=5.5cm, width=7.5cm]{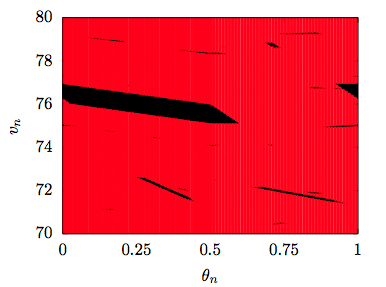} 
\caption{Basin of attraction for $a=19.0$, $\epsilon=0.9$. Black regions correspond to periodic solutions, and red regions correspond to sticking solutions. There is no chaotic behavior at these parameter values.}
\label{fig:basin1}
\end{figure}

If we scan the trapping region defined in section~\ref{sec:2.3} at parameter values where sticking and periodic solutions coexist, we may characterize the behavior fully for those particular values. Namely, we may distinguish the basins of attraction for the different attracting orbits. For the case $\epsilon=0.9$ $a=19.0$, a sticking solution and a multitude of periodic orbits coexist. In Fig.~\ref{fig:basin1} we plot a section of the trapping region which is characteristic of the general behavior for the whole range of initial conditions lying within the trapping region. 

\begin{figure}
\includegraphics[height=5.5cm, width=7.5cm]{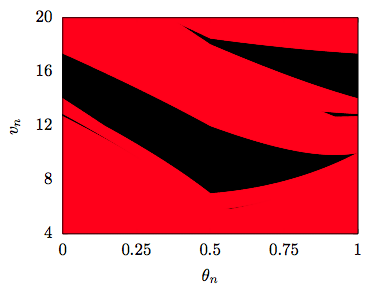}
\caption{Basin of attraction for the family of fixed points $(\theta_*,v_*)=(\theta_*,12)$ for $\epsilon=0.5$, $a=24.0$. Black and red regions correspond to fixed point and sticking solutions, respectively.}
\label{fig:basin2}
\end{figure}

As we lower $\epsilon$, the sticking solutions become on the whole more dominant for larger and larger ranges of $a$. However, in agreement with the earlier analytical investigations, there still exist special parameter values for which periodic solutions constitute a significant part of the trapping region. If we let $\epsilon=0.5$ and fix $a=24.0$, then we have a family of fixed point solutions at $\theta_*\in(0,\frac{1}{2}), v_*=12$ corresponding to the particle impacting with the wall once every wall oscillation. The basin of attraction for these orbits is given in Fig.~\ref{fig:basin2}. Clearly, the fixed point solutions present for precise parameter values constitute a large part of the trapping region, and are more dominant than periodic solutions present at other parameter values. 

\subsection{\label{sec:4.3}Global Behavior}
\begin{figure}
\includegraphics[height=5.5cm, width=7.5cm]{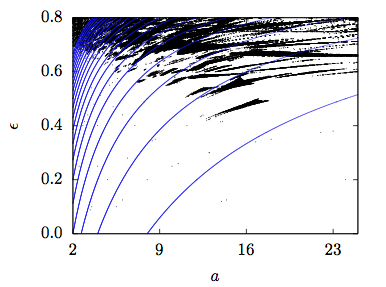}
\caption{Parameter space scan for periodic solutions. Note that above $\epsilon\sim 0.8$ periodic solutions are widespread. The blue curves are fixed point solutions.}
\centering
\label{fig:paramspace1}
\end{figure}

To gain a complete picture of the global dynamics, we scan the parameter space for different possible kinds of solutions. In our case, the prevalence of sticking solutions at all parameter values makes searching for such solutions trivial; hence, for the constant restitution model we scan the parameters for the existence of periodic (regular) solutions. In Fig.~\ref{fig:paramspace1} we display the parameter space, along with the fixed point solutions which were computed analytically; note that we have cut-off values of the restitution coefficient above $\sim 0.8$, as the existence of periodic solutions above this value is widespread. 

\section{\label{sec:5}Velocity-dependent model}
The dynamics for the velocity-dependent are richer and more complex due to the absence of sticking solutions, which as we have seen in the constant restitution model serve to destroy any long-term chaotic orbits. For this model we find the behavior may be classified as follows:

\begin{itemize}
\item Periodic solutions;
\item Aperiodic (chaotic) solutions;
\item Effective sticking solutions, where the particle experiences a large number of decaying collisions with the wall, and eventually matches the wall's velocity within the numerical precision of our calculations. Note that although the number of collisions is finite, we `effectively' observe the same end result of the process as in the case of true inelastic collapse.
\end{itemize}

\subsection{\label{sec:5.1}Bifurcation diagrams and typical trajectories}
\begin{figure}[h]
\subfloat[Subfigure 1 list of figures text][]{\includegraphics[height=5.5cm, width=7.5cm]{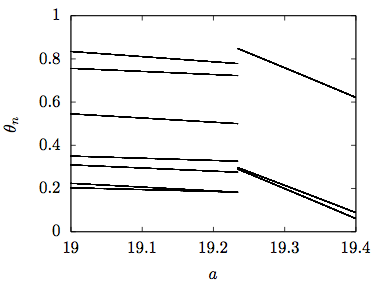}
\label{fig:bifurc_two}}
\qquad
\subfloat[Subfigure 2 list of figures text][]{\includegraphics[height=5.5cm, width=7.5cm]{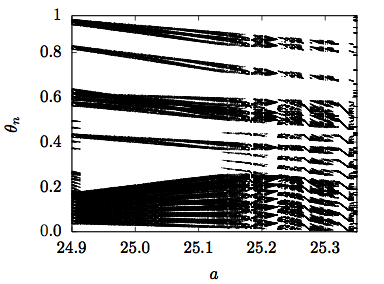}
\label{fig:bifurc_three}}
\caption{Bifurcation diagrams for the velocity-dependent model at $c\sim 1000$. (a) A period-$7$ orbit bifurcating to a period-$3$ orbit as one of the points in the orbit approaches $\theta_n=1/2$, and two points on the orbit converge. (b) periodic orbit bifurcating into a chaotic orbit as $a$ is decreased.}
\label{fig:bifurcationveldep}
\end{figure}

In considering the velocity-dependent model, we begin by fixing the constant in the restitution function and varying $a$. We observe drastically different behavior than in the constant restitution model, including a larger number of periodic orbits and more complex behavior. In Fig.~\ref{fig:bifurcationveldep} we display several typical ways an orbit can evolve as we vary $a$ at a constant value of $c$.  

In contrast to the constant restitution model, we observe what appears to be chaotic behavior resulting from the bifurcation of a periodic orbit into a chaotic structure. In section~\ref{sec:5.2} we compute the leading Lyapunov exponent for a range of $a$ values as well as the basin of attraction for this orbit, both of which confirm the chaotic nature of the dynamics at this parameter value. Furthermore, in section~\ref{sec:5.4} we survey the global behavior of the velocity-dependent model and find that chaotic behavior is widespread for a large range of parameter values. 

The existence of a positive leading average Lyapunov exponent is a commonly used indicator of chaos; hence we have computed this quantity for this particular range of $a$ values, the results of which confirm the chaotic nature of the orbit(s) at this parameter value. This indicates that nearby initial conditions exponentially diverge in time. We explore these attractors in greater depth in the next section. 

\subsection{\label{sec:5.2}Attractors}
\begin{figure}[h]
\subfloat[Subfigure 1 list of figures text][]{\includegraphics[scale=0.85]{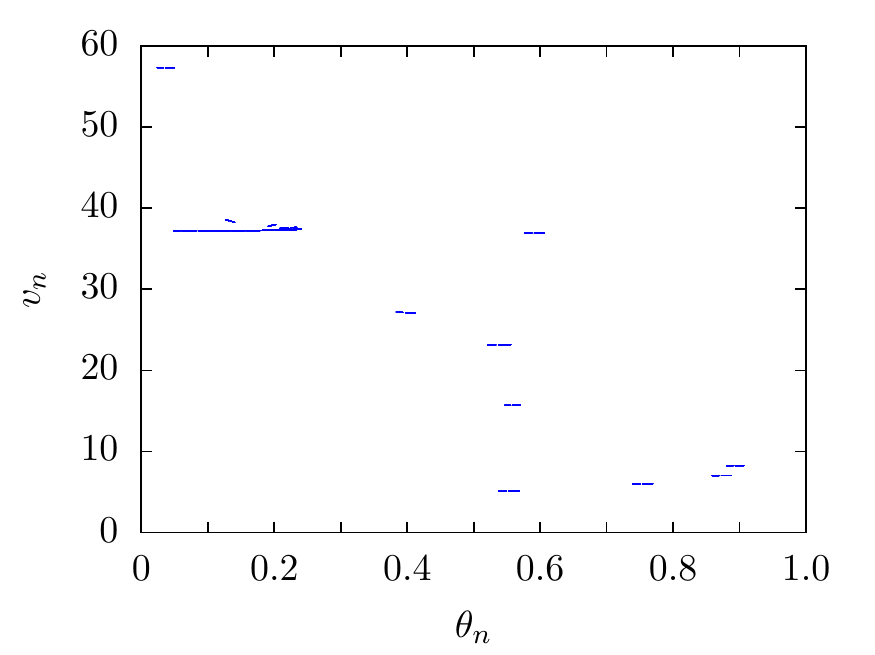}
\label{fig:attract_one}}
\qquad
\subfloat[Subfigure 2 list of figures text][]{\includegraphics[scale=0.85]{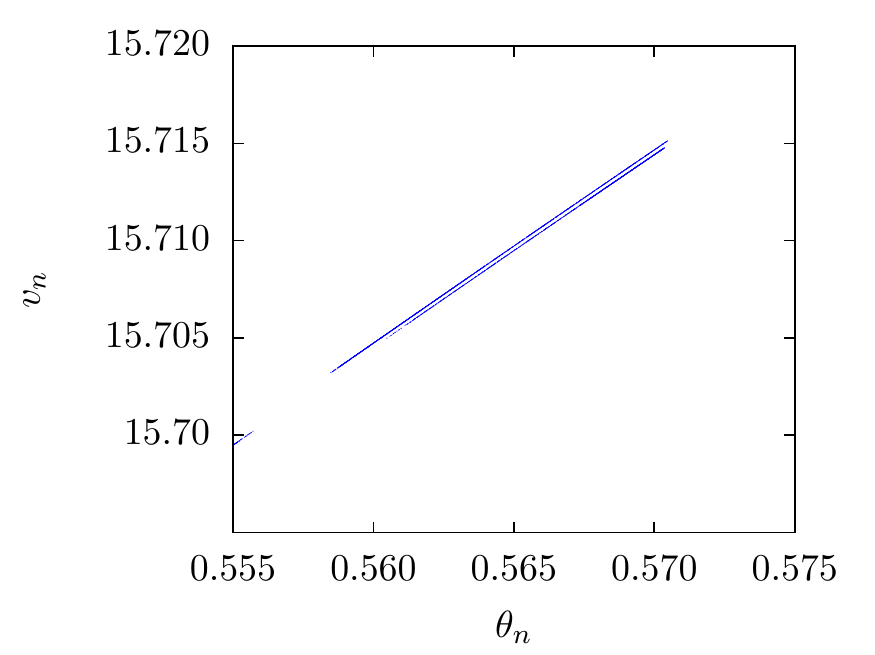}
\label{fig:attract_two}}
\qquad
\subfloat[Subfigure 2 list of figures text][]{\includegraphics[scale=0.85]{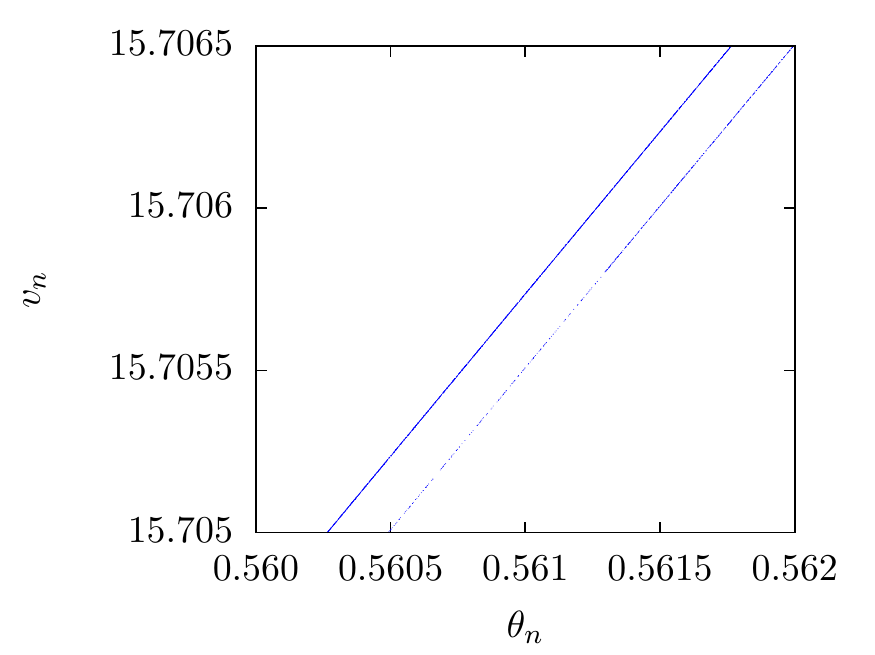}
\label{fig:attract_three}}
\caption{(a) Surface-of-section view of the strange attractor whose bifurcation origins are shown in Fig.~\ref{fig:bifurc_three} and discussed in section~\ref{sec:5.1}. (b)-(c) Zoomed-in views of the attractor; structure can be found at arbitrary scales.}
\label{fig:attractor}
\end{figure}

Despite the chaotic nature of the particle's trajectory at certain parameter values, we find that such orbits remain on a general structure for large numbers of collisions. In Fig.~\ref{fig:attractor} we give several views of the attracting orbit. Fig.~\ref{fig:attract_one} displays the full orbit, while Figs.~\ref{fig:attract_two}~and~\ref{fig:attract_three} show the structure present at smaller and smaller scales. We note that strange attractors observed in \cite{vogel2011} for the sinusoidal system are dramatically different in structure and origin than those found in our model. In the sinusoidally driven bouncer, chaotic attractors arise purely out of the classic period-doubling bifurcation cascades, whereas in our system routes to chaos are for the most part initiated by the discontinuity in the wall's motion. 

\subsection{\label{sec:5.3}Basin of attraction and Lyapunov exponent}
\begin{figure}[h]
\subfloat[Subfigure 1 list of figures text][]{\includegraphics[height=5.5cm, width=7.5cm]{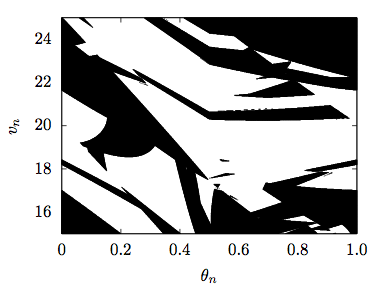}
\label{fig:basin_three}}
\qquad
\subfloat[Subfigure 2 list of figures text][]{\includegraphics[height=5.5cm, width=7.5cm]{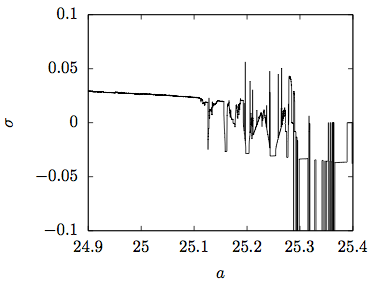}
\label{fig:lyapunov_one}}
\caption{(a) Black: basin of attraction corresponding to the velocity-dependent attractor for $a=25$, $c\sim 1000$. Zooming in on smaller regions of this reveals additional structure, supporting the chaotic nature of the orbit. (b) Leading Lyapunov exponent calculated for the attractor as $a$ is varied; the transitions from $\sigma<0$ to $\sigma>0$ correspond to the transitions on the bifurcation diagram in Fig.\ref{fig:bifurc_three}.}
\label{fig:basin3}
\end{figure}

We may gain an understanding of the influence of the chaotic attractor described in the previous section on the trapping region of phase space by computing the corresponding basin of attraction, which allows us to distinguish between the different kinds of possible long-term orbits. In Fig.~\ref{fig:basin_three} we display for $a=25$, $c\sim1000$ a representative section of the trapping region, showing initial conditions drawn to the regular and chaotic attractors for different $a$ values. The structure of the basin exhibits the property of continued structure upon magnification (i.e., a \textit{riddled} basin), lending further credence to the chaotic nature of the attractor. 

In addition to a riddled basin of attraction, the existence of a positive leading average Lyapunov exponent is a commonly used indicator of chaos; a positive value indicates that nearby trajectories diverge exponentially in time. In Fig.~\ref{fig:lyapunov_one} we plot the leading Lyapunov exponent $\sigma$ for the range of $a$ values corresponding to Fig.~\ref{fig:bifurc_three}. As expected, transitions from $\sigma<0$ to $\sigma>0$ correspond to points on the bifurcation diagram separating periodic and chaotic regions.

\subsection{\label{sec:5.4}Global behavior}
\begin{figure}[h]
\subfloat[Subfigure 1 list of figures text][]{\includegraphics[height=5.5cm, width=7.5cm]{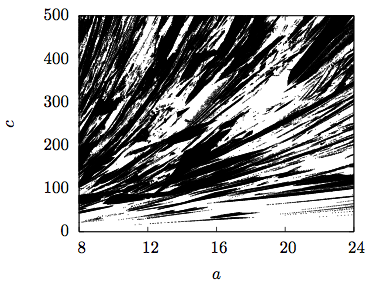}
\label{fig:paramscan}}
\qquad
\subfloat[Subfigure 2 list of figures text][]{\includegraphics[height=5.5cm, width=7.5cm]{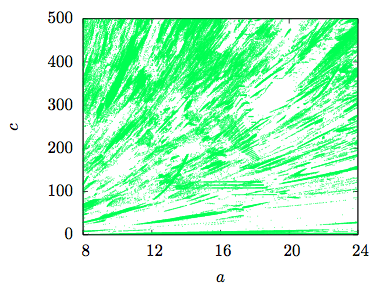}
\label{fig:paramscan2}}
\caption{Representative scans of the parameter space by solution type. (a) Black: periodic solutions. (b) Green: chaotic solutions.}
\label{fig:paramspace2}
\end{figure}

To understand the global solution structure, we scanned the parameter space for a wide range of initial conditions. We used a $10\times 10$ set of initial conditions in the trapping region defined in section~\ref{sec:2.3}. The results are shown in Fig.~\ref{fig:paramspace2}, illustrating the presence of periodic and chaotic solutions. For clarity we display only a representative portion of the parameter space, as in this model there is no precise upper bound for the parameter $c$ in the restitution function. We observe a similar global structure to the sinusoidal results seen in \cite{vogel2011}, as the reduced dominance of sticking/effective sticking solutions permits the existence of chaotic regions of the parameter space. 

\section{\label{sec:6}Conclusion}
The Fermi piston, and the closely related sinusoidal bouncer, are seminal models of chaotic driven dynamical systems. They are of general interest  because they exhibit a rich variety of phase space trajectories within a seemingly simple dynamical setting . In contrast with chaotic mappings, such as the standard map or Henon map, we can visualize the associated motion and, in some cases, observe it in the laboratory. Thus it is surprising that questions still remain about the nature of some of the possible motions. This may be due to subtleties in the numerical algorithms for simulating the dynamics that have yet to be completely resolved. Since impact models play an important roll in modeling more complex systems, such as one encounters in the study of granular media \cite{mcnamara1992,goldman1998}, a complete understanding of the associated behavior in simple models is essential.

To shed light on the true nature of the orbit geometry of a driven system with one space dimension, in this paper we have studied a piecewise linear bouncing ball system for three different collision models. In contrast to the widely studied sinusoidal model, we were able to derive the Poincar\'{e} map analytically; this was due to the piecewise linear nature of the wall's motion. We were also able to demonstrate the existence of fixed points and investigate their stability analytically. We showed the existence of higher order periodic orbits analytically and investigated their stability numerically. We found that the elastic model's phase space was, for the most part, dominated by chaotic accelerating orbits that surround stable quasiperiodic islands. However, an additional type of Fermi acceleration was found which is not present in the sinusoidal model; namely, we observed unbounded orbits which were almost periodic in the phase variable $\theta_n$. For inelastic collisions with a constant restitution coefficient, we computed the time required for inelastic collapse to occur analytically. In our numerical work on the constant restitution coefficient model we found that long-term trajectories are in general dominated by these sticking solutions. This is in close correspondence to the sinusoidal model. However, in contrast to the sinusoidal model we found a distinct lack of self-reanimating chaos i.e., all sticking solutions appear to be periodic. This gives credence to the idea that the self-reanimating chaos observed in the sinusoidal model could simply be a result of numerical errors.

Our velocity-dependent model has the advantage of eliminating sticking solutions which (in the constant restitution model) often prevent significant chaotic attractors from forming, allowing for a more diverse phase space. We demonstrated analytically that the time required for an infinite number of collisions diverges for our choice of restitution function. In our numerical simulations for the velocity-dependent model, we found periodic orbits to be prevalent, and observed the usual period-doubling and period-adding bifurcations expected for a system which is non-smooth. In addition, we found attractors where the trajectories stayed on a single structure for a large number of collisions. Together, the bifurcation processes which give birth to these attractors and a positive leading Lyapunov exponent suggest that the attractors are chaotic.

In future work, we would like to supplement our numerical results with analytical computations which explain the origins of the attractors in the velocity-dependent model. Investigating the extension of this work to a system of two particles colliding with each other and a piecewise linearly driven wall and the related problem of a particle falling in a driven wedge are possible future directions as well.

\begin{acknowledgments}
We would like to acknowledge the support of the REU program at TCU and the earlier work on this model carried out by Mathew Holtfrerich under this program. This work was supported by NSF grant PHY-1358770.
\end{acknowledgments}

\bibliography{piecewise_smooth_bouncer}

\end{document}